**Main Manuscript for**

# Quantifying urban and landfill methane emissions in the United States using TROPOMI satellite data

Xiaolin Wang[a], Daniel J. Jacob[a], Hannah Nesser[b], Nicholas Balasus[a], Lucas Estrada[a], Melissa Sulprizio[a], Daniel H. Cusworth[c], Tia R. Scarpelli[c], Zichong Chen[a], James D. East[a], Daniel J. Varon[a]

[a] School of Engineering and Applied Sciences, Harvard University, Cambridge, MA 02138, United States

[b] Jet Propulsion Laboratory, California Institute of Technology, Pasadena, CA 91011, United States

[c] Carbon Mapper, Pasadena, CA 91011, United States

*Corresponding author: Xiaolin Wang

**Email:** wangxi@g.harvard.edu

**Author Contributions:** XW and DJJ conceptualized the research. XW conducted the research with contributions from DJJ, ZC, JDE and DJV. HN contributed to the interpretation of inversion results. NB, LE and MS support the development of IMI. NB, DHC and TRS provided guidance on the landfill data analysis. XW and DJJ wrote the manuscript with input from all authors.

**Competing Interest Statement:** The authors declare that they have no conflict of interest.

**Classification:** Physical Sciences (major), Earth, Atmospheric, and Planetary Sciences (minor)

**Keywords:** Urban methane emissions; TROPOMI satellite; inverse modeling; landfills

**This PDF file includes:**

> Main Text
> Figures 1 to 4

**Abstract**

Urban areas are major sources of methane due to population needs for landfills, natural gas distribution, wastewater treatment, and residential combustion. Here we apply an inversion of TROPOMI satellite observations of atmospheric methane to quantify and attribute annual methane emissions at 12×12 km$^2$ resolution for 12 major US urban areas in 2022. The US Environmental Protection Agency Greenhouse Gas Inventory (EPA GHGI) is used as prior estimate. Our results indicate that the GHGI underestimates methane emissions by 80% on average for the 12 urban areas, with 22%–290% underestimations in most urban areas, except Los Angeles and Cincinnati where emissions are overestimated by 32%–37%. This is corroborated by independent surface-based observations in the Northeast Corridor and Los Angeles. Landfills are the principal cause of urban emission underestimates, with downstream gas activities contributing to a lesser extent than previously found. Examination of individual landfills other than in Los Angeles shows that emissions reported by facilities with gas collection and control systems to the Greenhouse Gas Reporting Program (GHGRP) and used in the GHGI are too low by a factor of 4 when using the prevailing recovery-first reporting method. This is because GHGRP-estimated gas collection efficiencies (average 70%, range 40–87%) are much higher than inferred from our work (average 38%, range 5–90%). Los Angeles landfills have much higher collection efficiencies (average 78% in GHGRP; 85% in our work) than elsewhere in the US, suggesting that operational practices there could help inform methane mitigation in other urban areas.

**Significance Statement**

High-resolution inversion of satellite observations for 12 large US urban areas shows that urban methane emissions are underestimated by 80% in the US Environmental Protection Agency (EPA) national inventory. The principal cause of this underestimate is landfills with gas collection and control systems that use too high collection efficiencies in their reported emission estimates. An exception is Los Angeles, where our results indicate much more efficient landfill gas collection than elsewhere. The low collection efficiencies observed in most cities reveal significant potential to reduce urban methane emissions through improved landfill management practices.

**Main Text**

**Introduction**

Rising atmospheric methane has contributed 30% of climate warming since pre-industrial times (1). Individual countries have committed to reduce their methane emissions through their Nationally Determined Contributions (NDCs) to the Paris Agreement and through the Global Methane Pledge to reduce anthropogenic methane emissions by 30% from 2020 levels by 2030 (2). Urban areas are major concentrated methane sources including natural gas distribution systems, landfills, and wastewater treatment plants. In the US, these sources account for one-fourth of total anthropogenic methane emissions according to the Greenhouse Gas Inventory (GHGI) of the US Environmental Protection Agency (EPA) (3). Individual cities increasingly develop their own goals to reduce greenhouse gas emissions (4, 5). Accurate quantification and attribution of urban methane emissions is essential to set emission reduction goals and monitor progress.

Urban emission inventories used for policymaking are generally constructed by bottom-up methods using emission factors applied to specific activities. Observations of atmospheric methane in surface air, from aircraft, and from satellites provide top-down information to evaluate these inventories using inverse methods. The TROPOspheric Monitoring Instrument (TROPOMI) on the Sentinel-5P satellite provides global daily observations of methane columns at 7×5.5 km² nadir pixel resolution (6). These and other atmospheric observations consistently show that US urban emissions are higher than the EPA GHGI (7, 8). Using ethane as an indicator for natural gas sources, aircraft-based studies have found that urban natural gas emissions are 2 to 10 times higher than the GHGI due to leakage underestimates and missing end-use emissions (9–11). Emissions from landfills and wastewater treatment plants, reported to the GHGI through the Greenhouse Gas Reporting Program (GHGRP), have also been found to be biased low compared to atmospheric observations (7, 12–16).

Here, we use an inversion of TROPOMI observations with Bayesian optimization to estimate emissions from US urban areas at 12-km resolution. TROPOMI observes total atmospheric methane, averaging over temporal variability and including contributions from both area and point sources. Optimizing emissions at 12-km resolution enables better separation between source types (sectors) than previous TROPOMI inversions conducted at coarser resolution. We apply the Integrated Methane Inversion (IMI), which determines an optimal (posterior) estimate of emissions by analytical minimization of a Bayesian cost function including prior emission estimates from the GHGI (17, 18). The IMI analytical solution includes detailed error characterization for the posterior estimate through the posterior error covariance matrix. We present results for 12 major urban areas in the U.S in 2022, separating contributions from different sectors, and evaluate these results with independent surface and aircraft observations.

Our analysis resolves emissions from individual major landfills which we compare to the bottom-up models used by these landfills to report their emissions to the GHGRP.

## Results and Discussion

### Urban methane emissions

We focus on 12 large urban areas in the US as shown in Fig. 1. Urban boundaries follow definitions from the US Census Bureau as the densely settled census blocks that meet minimum housing unit or population density thresholds (19). We selected the 10 most populous urban areas (by rank New York, Los Angeles, Chicago, Dallas, Houston, Miami, Philadelphia, Atlanta, Washington, Boston). Cincinnati and Detroit are also included because of high emissions previously inferred by Nesser et al. (7) in their TROPOMI inversion. These 12 urban areas account for 27% of the Contiguous United States (CONUS) population but only 1.5% of the land area. Their high population density and concentrated emissions make them key targets for evaluating methane emissions and developing mitigation strategies.

Fig. 1 shows mean TROPOMI observations of dry-column methane mixing ratios ($XCH_4$) for 2022 in the 12 urban areas. We use the blended TROPOMI+GOSAT satellite product (20) designed to correct retrieval artifacts in the operational TROPOMI data by application of machine learning trained on differences with data from the Greenhouse Gases Observing Satellite (GOSAT). We further filter out water and mixed water-land pixels, which have residual artifacts (20). GOSAT uses the $CO_2$ proxy method to avoid surface reflectivity and aerosol scattering artifacts (21), but this can induce bias in environments with high $CO_2$ variability. We find no indication of such bias for the urban areas examined here based on comparison with the operational TROPOMI product, which uses a full-physics retrieval independent of $CO_2$ concentrations (Fig. S1).

The observations in Fig, 1 show varying levels of methane enhancements in the urban areas from which we can infer emissions. We quantify annual methane emissions in each urban area for 2022 by inversion of the TROPOMI observations over a 3°×4° domain including the urban area and its surroundings (hereafter called the extended urban domain). Fig. 2 illustrates the inversion procedure for New York City and its extended urban area. Urban emission totals are estimated as the sum of grid cells within and intersecting the urban boundaries (thick black lines). The same Figures for all other urban areas are in the *SI* (Fig. S2–S3). Fig. 2A to Fig. 2D shows the prior emissions separated by sectors with anthropogenic sources from the gridded GHGI at 0.1°×0.1° resolution including GHGRP facility-level source information (22). Wetland emissions are from WetCHARTs v1.3.1 at 0.5° × 0.5° resolution (23). The GEOS-Chem model used in the IMI is applied here at 0.125°×0.15625° (≈ 12 × 12 km$^2$) spatial resolution, and the optimization of emissions is performed at that same resolution.

We perform two sets of inversions using either the blended TROPOMI+GOSAT product or the operational TROPOMI product v02.04.00, with the blended product as our base inversion. Prior estimates of boundary conditions for the inversions are from smoothed TROPOMI observations following IMI procedure so as to be consistent with the observations used. We optimize these boundary conditions annually on the four cardinal domain edges as part of the inversion. We conduct an ensemble of inversions to determine the sensitivity to inversion parameters and report the ensemble mean as the posterior estimate (Fig. 2E). The uncertainty range in posterior estimates is given as the spread of the inversion ensemble (Fig. 2G). The ability of the inversion to separate emissions from different sectors is given by the posterior error correlation matrix (Fig. 2H). See *Methods* for further details of the inversion procedure.

For independent evaluation of the inversion results, we compare GEOS-Chem simulations driven by either prior or posterior emissions with tower-based methane observations from 18 sites in the Northeast Corridor network (24, 25) covering New York, Washington, and Philadelphia. We also use ground-based column observations in Los Angeles from the two sites of the Total Carbon Column Observing Network (TCCON) (26). As shown in Fig. 2F, S4, and Table S1, the fit to these independent observations is improved when using posterior emissions.

We define urban emissions as those from population-driven activities including downstream gas, landfills, wastewater treatment, and other small sources such as stationary combustion. Downstream gas includes distribution from city gates to consumers and post-meter end use. We separate the contributions of individual sectors in our posterior estimates based on the relative contributions in the prior estimates for each 12-km grid cell. In the gridded GHGI, landfill emissions are mapped based on GHGRP facility coordinates, showing clear separation from other sectors with posterior error correlations below 0.35 (Fig. 2H and S3). Emissions from downstream gas activities, wastewater treatment in septic systems, and stationary combustion (categorized as 'other anthropogenic sources' in the GHGI) are all allocated based on population and this results in error correlations ranging from 0.45 to 0.87. Emissions from wastewater treatment and stationary combustion are relatively small compared to landfills.

Fig. 3 shows emissions by sector for the 12 urban areas. Total emissions from the 12 urban areas in the GHGI are 1.0 Tg a$^{-1}$ (59% landfills, 25% downstream gas, 9% wastewater, 7% other anthropogenic). Our posterior estimate is 80% higher at 1.8 Gg a$^{-1}$ (62% landfills, 23% downstream gas, 8% wastewater, 7% other anthropogenic). We find large differences between urban areas. Landfills and downstream gas dominate everywhere, but the contribution from landfills in the posterior estimate varies from 28% in Washington to 84% in Miami while the contribution from downstream gas varies from 14% in Atlanta to 44% in New York and Boston. Emission corrections for individual urban areas vary from 32%–37% overestimates in Los Angeles and Cincinnati to a 290% underestimate in Houston. These emission corrections are primarily attributed to landfills. Emissions from downstream natural gas are underestimated mainly in Houston, Detroit, New York, and Washington.

Our posterior estimates are highest in Houston, Detroit, Dallas, and Atlanta, despite their smaller populations than New York and Los Angeles. Per capita emissions in New York and Los Angeles are low relative to other cities. We find that variations in per capita emissions are largely driven by landfill emissions. As shown in Fig. 3A, landfill emissions per capita per year range from 36 Gg per million in Detroit to 3 Gg per million in New York, Boston, and Washington. Urban landfill emissions correlate more strongly with total waste-in-place (R=0.58) than with population (R=-0.16) (Fig. S5). According to GHGRP landfill data, New York, Boston, and Washington report lower per capita landfill waste than other cities. These urban areas tend to have fewer active landfills and lower disposal mass per landfill in 2022, largely due to zero-waste initiatives promoting waste diversion from landfills to waste-to-energy incineration facilities outside urban areas, along with increased recycling and composting efforts. In contrast, downstream gas emissions are strongly correlated with urban population (R=0.48; R=0.79 excluding Houston) and gas usage (R=0.40; R=0.78 excluding Houston) (Fig. S5). Gas emission per capita per year ranges from 2.1 Gg to 5.6 Gg per million across urban areas, except in Houston and Detroit where it is much higher.

Our findings identify landfills as the primary source of urban emission biases, consistent with remote sensing studies on the national scale (7) and for numerous landfill sites (12, 13, 15, 27). However, in-situ and aircraft-based studies using ethane as a tracer have attributed approximately 80% of methane emissions to natural gas in Boston, Philadelphia, Washington, Los Angeles, and New York (9, 10, 28), much higher than our estimates of 22% to 44% in these areas. These studies have tended to emphasize gas-dominated plumes from city cores. They focused less on landfills outside city cores. When limiting our analysis to core urban areas (60% of the population), the estimated natural gas contribution rises to 33%–55%. According to the GHGI, emissions from gas distribution in these urban areas declined by 10%–15% between 2012 and 2020. This reduction is largely due to the replacement of leaky cast iron pipelines with plastic alternatives, which may also contribute to the lower gas emissions in our estimation.

Inversion results using the TROPOMI v02.04.00 operational product (29) are higher than the blended TROPOMI+GOSAT product (20) over the Northeast Corridor and Miami (Fig 3B). The differences between the two products are most pronounced over dark surfaces (mostly wetlands), where GOSAT has better observational capability than TROPOMI and where the operational TROPOMI product shows higher urban methane enhancements (Fig. S6). This explains the higher inversion results in the Northeast Corridor and Miami.

Fig. 3B also compares our results with 16 top-down studies for different urban areas published since 2015. These may use different definitions of urban areas and generally include contributions from all sources. Beyond urban-specific sources, emissions in Dallas and Houston include nearby oil and gas operations, while emissions in the Northeast Corridor and Miami have important contributions from wetlands. Our inversion results are most immediately comparable to Nesser et al. (7), who performed an analytical inversion of TROPOMI observations over CONUS using an earlier version of the operational TROPOMI product and the same definition of urban areas as used here. Our results agree in Los Angeles, Dallas, and Detroit but we tend to be lower in other cities. The discrepancies may be largely driven by differences between TROPOMI products. Our inversion using the operational TROPOMI data aligns more closely with Nesser et al. (7) for Atlanta, New York, and Chicago.

Urban emissions in other studies are typically derived from aircraft and in-situ measurements. Most apply inverse methods with the GHGI as prior estimate, whereas Karion et al. (30) and Pitt et al. (31, 32) developed finer-resolution city inventories for Washington and New York, respectively. Plant et al. (31) applied aircraft and TROPOMI $CH_4$:CO or $CH_4$:$CO_2$ ratios combined with CO and $CO_2$ inventories to estimate emissions. Direct comparisons are complicated by differences in urban area definitions and reporting periods, as summarized in Table S2 and discussed in Nesser et al. (7). These previous studies generally derive higher emissions than ours, though independent evaluation with tower and TCCON data is consistent with our estimates. In Los Angeles at least, part of the difference is attributable to declining emissions over time (-7% per year) (7, 33), and the use of a broader domain in other studies that covers the South Coast Air Basin, where total emissions are about 40% higher than within our urban boundaries.

**Emissions from individual landfills**

In the GHGI, landfill emissions are based on GHGRP data submitted by individual facilities. The 12×12 $km^2$ resolution of our inversion enables us to quantify annual emissions from large individual landfills and compare to GHGRP reports. We apply this to landfills not only in the urban areas but in the extended 3°×4° extended urban domains of Fig. 1.

The GHGRP requires landfill operators to report methane emissions ($E_{CH4}$) using a bottom-up model:

$$E_{CH4} = (G_{CH4} - R) \times (1 - OX) + D \quad (1)$$

where $G_{CH4}$ is the net methane generation from the anaerobic decomposition of organic waste and $R$ represents the methane recovered by the gas collection and control system (GCCS). For landfills without a GCCS, $R$ is zero. $OX$ is the fraction of methane oxidized by the landfill cover and ranges from 0% to 35% depending on cover type. $D$ accounts for the recovered methane that ends up being emitted due to leaks or incomplete destruction.

Landfill operators with a GCCS must calculate $G_{CH4}$ using two alternative approaches, recovery-first and generation-first, and select one as the reported value under GHGRP (34). In the recovery-first method, $G_{CH4}$ is back-calculated from the measured recovered methane quantity ($R$) and a collection efficiency ($CE$) as:

$$G_{CH4} = \frac{1}{CE} \times \frac{R}{f_{Rec}} \quad (2)$$

where $f_{Rec}$ is the fraction of time that the recovery system operates. Uncertainty lies in the estimate of $CE$. In the generation-first method, $G_{CH4}$ is estimated using first-order decay rates applied to the degradable organic carbon content in total waste. This relates landfill emissions to waste-in-place but is subject to uncertainties associated with the waste composition and decay rates (35, 36). For landfills without a GCCS, only the generation-first model is applied. 75% of 1123 landfills in the US reporting to the GHGRP in 2022 were equipped with a GCCS, and 74% of those chose the recovery-first method.

Our study encompasses 381 municipal solid waste (MSW) landfills across the 12 extended urban domains of Figure 1. We limit our analysis to 53 landfills (43 with GCCS and 10 without) for which we have adequate sensitivity and which account for >80% of prior emissions in their 0.1°×0.1° GHGI grid cell. Fig. 4A compares our posterior emissions with 2022 GHGRP bottom-up model values for individual landfills, including 9 landfills in Los Angeles and 44 landfills in other urban areas. The recovery-first method is a factor of 3.7 lower than our estimate based on the regression slope (R = 0.29). The generation-first method is 30% lower than our estimate and captures more of the variability (R = 0.41). Los Angeles is an exception where we estimate lower emissions than either GHGRP method.

The gas collection efficiency CE (Eq. (2)) is a major uncertainty in the recovery-first method because it cannot be measured directly (35). The GHGRP recommends collection efficiency estimates of 0% for working faces without active gas collection (cover area $A2$); 60% for daily cover (*A3*), 75% for intermediate cover (*A4*), and 95% for final cover (*A5*) (34). *A1* is the landfill area without waste in place and is not used in the calculation. An area-weighted average collection efficiency is then calculated for the entire landfill as *CE*= (0.6*A3*+0.75*A4*+0.95*A5*)/$\sum_{i} A_i$ where the sum is over areas *A2-A5*. A default efficiency of 75% is to be used if no cover type information is available.

For the 44 landfills in our analysis excluding Los Angeles, GHGRP estimates gas collection efficiencies between 40% and 87% with an average of 70% (Fig. 4B). Using our posterior estimates, we recalculated collection efficiencies based on GHGRP-reported data for *R, f_{rec}, OX,* and *D* in Eq. (1) and (2). We find the collection efficiency in landfills other than Los Angeles to range from 5% to 90% with a mean of 38%. Overestimation of collection efficiency has been previously identified as a key factor contributing to underestimated landfill emissions in the GHGRP (7). To address this issue, EPA proposed amendments to reduce assumed collection efficiencies by 10% starting in January 2025. Our findings indicate that this adjustment is still insufficient and an average reduction of approximately 30% is needed. Consistent with our findings, Giordano et al. (37) inferred collection efficiencies of 41% for daily cover, 69% for intermediate cover, and 71% for final cover from in-situ measurements at 154 landfills in the US, Europe, and Canada. But we also find a very large variability in collection efficiencies that cannot be captured by the simple *CE* model.

The nine Los Angeles landfills in our analysis show high collection efficiencies averaging 85% (78–92%). This reflects California's effort to reduce landfill emissions through enhanced gas collection and control measures. For example, Sunshine Canyon landfill underwent substantial infrastructure upgrades since 2017 (38) through the addition of gas wells, installation of pumps, and replacement of the final soil cover with closure turf and vegetative cover. These measures effectively enhanced gas well performance, resulting in a collection efficiency increase to 92% in 2022 in our analysis. Our collection efficiency estimates also depend on the assumed OX values. While values of 0.1–0.35 are commonly used in Los Angeles, they likely reflect convention rather than site-specific measurements (39), and higher OX values would yield lower CE estimates under Eqs. (1) and (2). If the mean collection efficiency for US landfills in our analysis were raised to the level achieved in Los Angeles (from 38% to 85%), urban landfill methane emissions would be reduced by a factor of four.

The variability in collection efficiency is further influenced by site-specific practices related to cover materials and gas capture systems. For example, methane leaks can result from leachate system failures caused by excessive precipitation, maintenance deficiencies, or cracked covers. Such factors reduce collection efficiency but are not accounted for in the GHGRP models.

Disturbances from GCCS installation could also lead to high emissions (7, 12). Airborne point-source detections reveal that a large fraction of emissions associated with landfills is at the working face (27). Working face areas should be classified as *A2* (0% collection efficiency) but 19 landfills in our analysis reported no *A2* areas.

In summary, we quantified methane emissions by sector in 12 US urban areas using a high-resolution (12×12km$^2$) inversion of TROPOMI satellite observations. Compared to the GHGI national emission inventory, we find that urban methane emissions are underestimated by up to 290% in Houston and overestimated by 32%–37% in Los Angeles and Cincinnati. Landfills account for 62% of urban emissions and represent the primary source of underestimation. Downstream gas emissions (23%) are also underestimated in some cities but not as much as previously reported. By investigation of emissions from individual landfills, we find that Los Angeles landfills have much higher gas collection efficiencies (averaging 85%) than other cities (averaging 38%), apparently reflecting better equipment and control measures. Urban landfill emissions in the US could be reduced by a factor of four if collection efficiencies were increased to Los Angeles values.

**Materials and Methods**

We inferred methane emissions for 2022 in 12 US urban areas by inversion of TROPOMI satellite observations of atmospheric methane. The extent of the 12 urban areas is defined using the US Census Topographically Integrated Geographic Encoding and Referencing (TIGER) database (19). For each urban area, we infer emissions over a 3° × 4° (latitude × longitude) domain, including the urban area and its surroundings. This extended urban domain accounts for emissions from nearby landfills potentially serving the urban areas. We conduct our inversions using the IMI v2.0 (17, 18), with the GEOS-Chem chemical transport model to simulate methane concentrations and their sensitivities to emissions. The IMI performs analytical inversions to infer posterior estimates of emissions including closed-form error characterization. This *Methods* section describes the different components of the inversion: (1) the satellite observations from TROPOMI, (2) the GEOS-Chem model at 12×12 km$^2$ resolution, (3) the inversion framework in the IMI, (4) the methodology for attributing posterior emissions to different sectors and to individual landfills for comparison with GHGRP landfill data.

**TROPOMI satellite observations**

TROPOMI is a passive grating imaging spectrometer launched onboard the low Earth orbit Sentinel-5 Precursor satellite in October 2017. It provides daily global observations of atmospheric methane with a nadir pixel resolution of 5.5 × 7 km² and a local overpass time around 13:30. The operational product of XCH$_4$ is retrieved under clear-sky conditions from the shortwave infrared (SWIR) solar backscatter radiation in the 2.3 µm methane absorption band (6, 29). Balasus et al. (20) used machine learning to correct retrieval artifacts in the operational TROPOMI XCH$_4$ using the more precise (but much sparser) CO$_2$ proxy retrievals from the GOSAT satellite. We use this blended TROPOMI+GOSAT product for our base inversion and also use the operational TROPOMI product v02.04.00 (40) in a sensitivity inversion. We apply filtering criteria recommended by Balasus et al. (20) to exclude full-water pixels (surface_classification = 1), coastal pixels influenced by partial water surfaces (surface_classification = 3), and inland water pixels with poor spectral fits (surface_classification = 2 and SWIR chi-square > 20,000). For the operational TROPOMI product, we apply filters to exclude retrievals with QA values ≤ 0.5 and full-water pixels following Estrada et al. (17).

**GEOS-Chem forward model**

We use the GEOS-Chem 14.4.1 (https://doi.org/10.5281/zenodo.12584192) chemical transport model to simulate methane concentrations and their sensitivity to emissions. GEOS-Chem is driven by Goddard Earth Observation System – Forward Processing (GEOS-FP) assimilated meteorological data from the NASA Global Modeling and Assimilation Office (GMAO). The simulation uses the nested GEOS-Chem Classic version of the model over 3°×4° domains at a horizontal resolution of 0.125°×0.15625° (≈ 12×12 km$^2$) with 47 vertical levels. Boundary conditions are from smoothed TROPOMI concentrations (10°×12.5° spatially and - 15 days temporally) applied to a global GEOS-Chem simulation with 2°×2.5° resolution, and including zonal mean corrections over oceans, as described in Estrada et al. (17). This approach minimizes biases in the boundary conditions (but they are still optimized in the inversion, see below) and ensures that the inversions of blended and operational TROPOMI products each use their own consistent boundary conditions.

The 12-km resolution is a new GEOS-Chem capability enabled by the archival of native-resolution c720 cubed-sphere advection data in the GEOS-FP product and regridded here to 12×12 km$^2$ for use in the nested GEOS-Chem Classic model. This advection archive includes mass fluxes, surface pressure, and specific humidity at hourly temporal resolution. The mass fluxes are used to calculate horizontal wind vectors for input into GEOS-Chem Classic. Other meteorological variables including convective mass fluxes and vertical mixing depths are from the standard GEOS-FP product at 0.25° × 0.3125° (≈25×25 km$^2$) resolution and are regridded to 12×12 km$^2$.

**IMI inversion framework**

The IMI optimizes a gridded emission state vector $x$ following the Bayesian analytical inversion technique (41). For each extended urban area, the state vector $x$ is defined as the annual emissions at the native GEOS-Chem grid resolution (12×12 km$^2$) over the 3° × 4° domain, resulting in approximately 600 emission elements to be optimized for each area plus 4 additional elements for boundary conditions (north, south, east, west). We assume a lognormal error distribution for prior emissions to capture the heavy tail of the emission distribution and maintain positive solutions, while the prior error of boundary conditions follows a normal distribution. In the lognormal space, we optimize $x' = ln(x)$ instead of $x$ (42). The IMI solves the optimal estimate of $x'$ by minimizing the Bayesian cost function $J(x')$:

$$J(x') = (x' - x'_a)^T S'^{-1}_a (x' - x'_a) + \gamma(y - K'x')^T S_o^{-1}(y - K'x') \quad (1)$$

where $x'_a = \ln(x_a)$ and $x_a$ is the median of prior estimates for $x$. The posterior estimate $\hat{x}'$ is obtained by solving for the minimum of the cost function (d$J$/d$x'$=**0**) iteratively as described by Estrada et al. (17) and Hancock et al. (43). $S'_a$ is the prior error covariance matrix assuming diagonal elements $s'_a = ln^2(\sigma_g)$ for emissions, and $s'_a = \sigma_b^2$ for boundary conditions. Here $\sigma_g$ is the geometric error standard deviation for emissions and $\sigma_b$ is the error standard deviation for boundary conditions. $y$ is the ensemble of TROPOMI super-observations obtained by averaging the individual retrievals for a given orbit (observation day) over GEOS-Chem 12×12 km² grid cells (44). The observational error covariance matrix $S_o$ is constructed using the residual error method and accounts for the error covariances when averaging individual observations into super-observations. $K'$ is the Jacobian matrix representing the sensitivity of $y$ to $x'$ calculated by perturbing the emission elements in GEOS-Chem model. γ is a regularization factor to prevent overfitting to observations. Overfitting is implied if the state component of the cost function $J_A(\hat{x}') = (\hat{x}' - x'_a)^T S'^{-1}_a (\hat{x}' - x'_a)$ is much larger than the expected value of $n \pm \sqrt{2n}$ (45). In our case, $J_A$ is lower or within the expected ranges, so we set γ to 1.

The posterior error covariance matrix $\hat{S}'$, characterizing uncertainty in $\hat{x}'$, is given by:

$$\hat{S}' = \left(\gamma K'^T S_o^{-1} K' + S'^{-1}_a\right)^{-1} \quad (2)$$

Uncertainty in the choice of inversion parameters adds to the error in posterior estimates. To address this, we generate an ensemble of six inversions by varying (1) the geometric error standard deviation $\sigma_g$ =2, 3, or 4, (2) the error standard deviation for boundary conditions $\sigma_b$ =5, 10 ppb. We report the mean value of the inversion ensemble and use the range between the minimum and maximum to represent the uncertainties.

The averaging kernel matrix is defined as $A = \frac{d\hat{x}}{dx}$ and represents the sensitivity of the posterior emissions to the true value $x$. The averaging kernel matrix in log space $A' = \frac{\partial \hat{x}'}{\partial x'} = I_n - \hat{S}' S'^{-1}_a$ has elements $\frac{\partial ln(\hat{x}_i)}{\partial ln(x_j)} = \frac{x_j}{\hat{x}_i} \frac{\partial \hat{x}_i}{\partial x_j}$. However, $A'$ does not directly correspond to the desired $A = \frac{d\hat{x}}{dx}$ and one cannot infer $A$ from $A'$ because the true value $x$ is unknown. Instead, we construct $A = \frac{d\hat{x}}{dx}$ through an additional inversion assuming normal error distributions for $x$ with a 50% error standard deviation in the prior emissions, using a γ value of 1. The trace of $A$ defines the degrees of freedom for signal (DOFS), which represents the number of independent pieces of information on the emissions from the observations. We find DOFS ranging from 1.4 in Cincinnati to 17 in Dallas, with an average value of 7.9 for the 12 urban areas.

**Attributing posterior emissions to individual sectors and landfills**

The posterior emissions $\hat{x}$ obtained by the inversion in $n$ 12×12 km² grid cells are aggregated to different sectors in each urban area (black contours in Fig. 1) by applying a transformation matrix **W**. The relative contributions from the $q$ individual sectors to the total posterior emissions in each grid cell are taken from the prior estimates to produce a matrix **W** ($q$×$n$). We then apply **W** to transform the posterior state vector ($n$×1) to the reduced state vector $\hat{x}_{red}$ ($q$×1) and posterior error covariance $\hat{S}_{red}$ ($q$×$q$) as:

$$\hat{x}_{red} = W\hat{x} \quad (5)$$

$$\hat{S}_{red} = W\hat{S}W^T \quad (6)$$

Here $\hat{x}_{red}$ represents the aggregated emissions of different sectors in a given urban area. The error correlation matrix between sectors as shown in Fig. 2H has elements $R_{ij} = \frac{s_{ij}}{\sqrt{s_{ii} \cdot s_{jj}}}$, where $s_{ij}$ are the elements in $\hat{S}_{red}$.

We compare our results for landfill emissions at the facility level to GHGRP data accessed via the Facility Level Information on GreenHouse gases Tool (FLIGHT) (46). Our inversion results, originally on the 0.125°×0.15625° grid, are regridded to 0.1°×0.1° resolution to enable direct comparison with the GHGRP. Our inversions cover 381 GHGRP landfills across 12 extended urban domains (3° × 4° domains), from which we select individual landfills for which the inversion provides sufficient information (grid cell averaging kernel sensitivity > 0.1) and separates them from other sources (only one landfill per grid cell accounting for over 80% of prior emissions). This yields 53 landfills for comparison with the GHGRP.

**Data, Materials, and Software Availability**

The IMI source code and documentation is available at https://imi.seas.harvard.edu/. The TROPOMI methane data are available on the Amazon Web Services (AWS) cloud at https://registry.opendata.aws/sentinel5p/ (last access: 28 March 2024). The blended TROPOMI+GOSAT data are available at https://registry.opendata.aws/blended-tropomi-gosat-methane/ (20). The TCCON observation dataset are available at https://doi.org/10.14291/TCCON.GGG2020 (last access: 03 September 2024). Gridded prior and posterior emissions from Karion et al. (30) are available at https://doi.org/10.18434/mds2-3720. Our urban emission data from the TROPOMI inversion and urban shapefiles are archived at Zenodo (https://doi.org/10.5281/zenodo.15377282) and is currently under embargo. The dataset will be made publicly available upon publication.

**Acknowledgments.** We thank Anna Karion (National Institute of Standards and Technology) for providing methane observation data from the Northeast Corridor tower network (25) and gridded emission data for Washington (47). This work was funded by the Harvard Methane Initiative and by the NASA Carbon Monitoring System.

46. US Environmental Protection Agency, Facility Level Information on GreenHouse gases Tool (FLIGHT). Available at: https://ghgdata.epa.gov/ghgp/main.do [Accessed 15 April 2024].

47. A. Karion, *et al.*, Methane emissions estimates and related data sets for Washington DC and Baltimore, MD urban areas. National Institute of Standards and Technology. https://doi.org/10.18434/MDS2-3720. Deposited 7 February 2025.

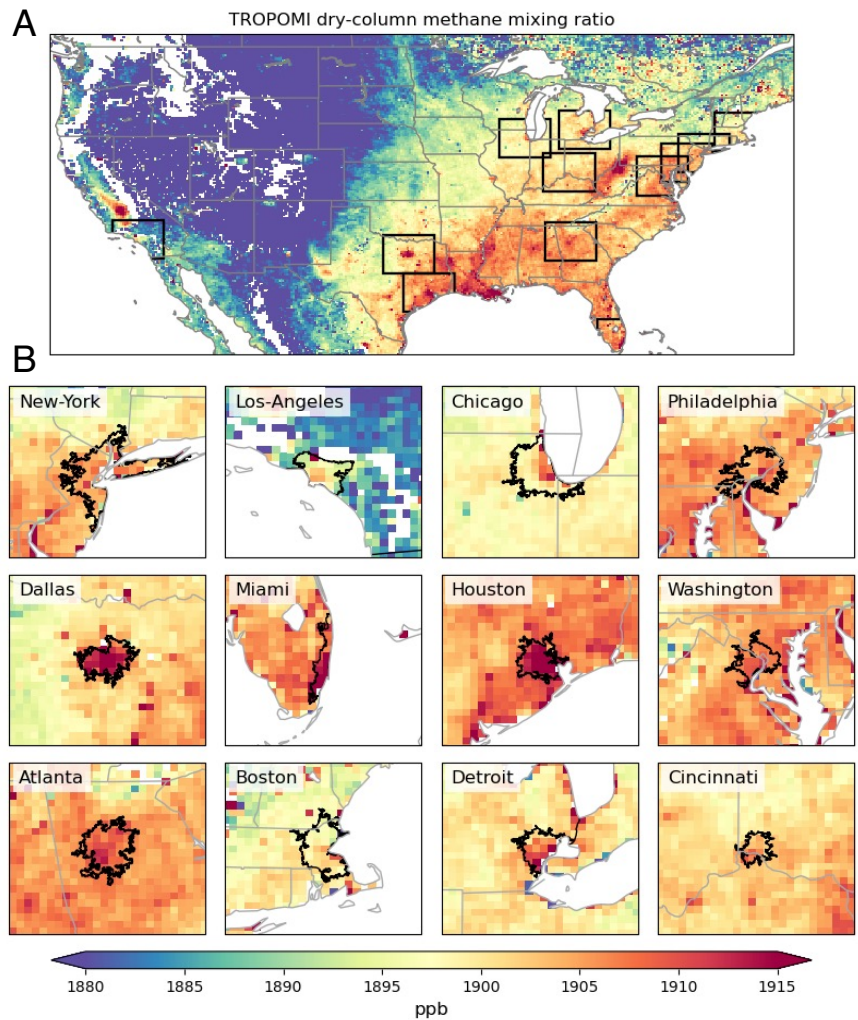

**Fig. 1.** TROPOMI satellite observations of dry-column methane mixing ratios (XCH$_4$) over US urban areas. Values are 2022 annual means from the blended TROPOMI+GOSAT product (20) on the 0.125° ×0.15625° (≈12 × 12 km$^2$) inversion grid. White areas have no observations. Black rectangles outline the 3°×4° simulation domains used in the inversions for the 12 urban areas. These domains are shown in the lower panels with urban boundaries from the US Census Bureau (19).

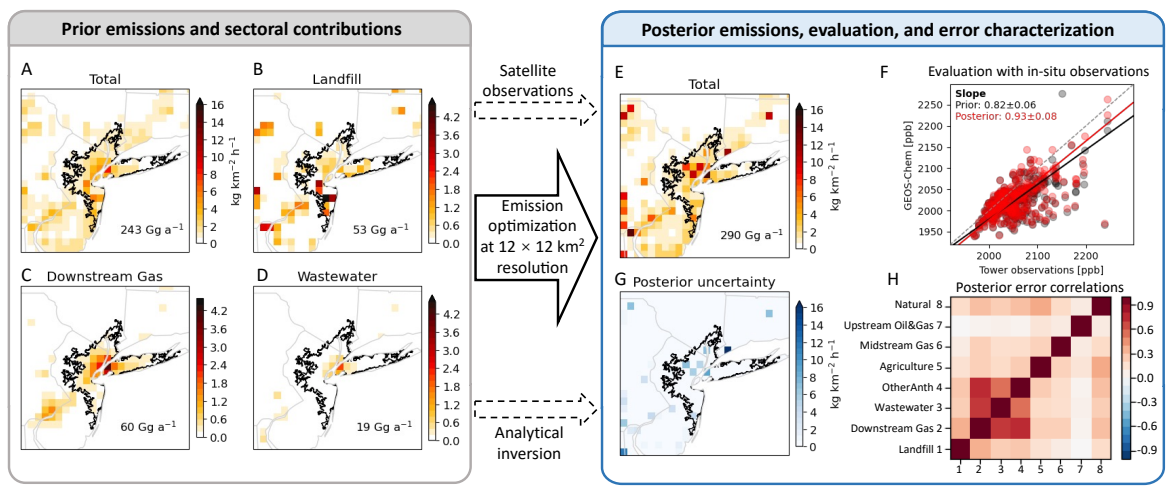

**Fig. 2.** Optimization of methane emissions in New York City with 12 × 12 km$^2$ resolution by inversion of TROPOMI satellite observations for 2022. Panel A shows prior emission estimates for the inversion including both anthropogenic emissions from the gridded Greenhouse Gas Inventory (GHGI) (22) and natural sources. Also shown are contributions from urban sectors (B, C, D). Urban area totals are shown inset as the sum of grid cells within and intersecting the urban boundaries (thick black lines). Panel E shows the optimized posterior emissions from the mean of our inversion ensemble, panel G shows the uncertainty in posterior emissions as the spread of the inversion ensemble, and panel H shows the posterior error correlation matrix for different sectors. Also shown in panel F is an independent evaluation of the inversion results by comparison of the GEOS-Chem model simulation with either prior or posterior emissions to the Northeast Corridor tower-based atmospheric concentration observing network (24, 25), which was not used in the inversion (F). Slopes are from a reduced-major-axis regression and error standard deviations are from bootstrap resampling.

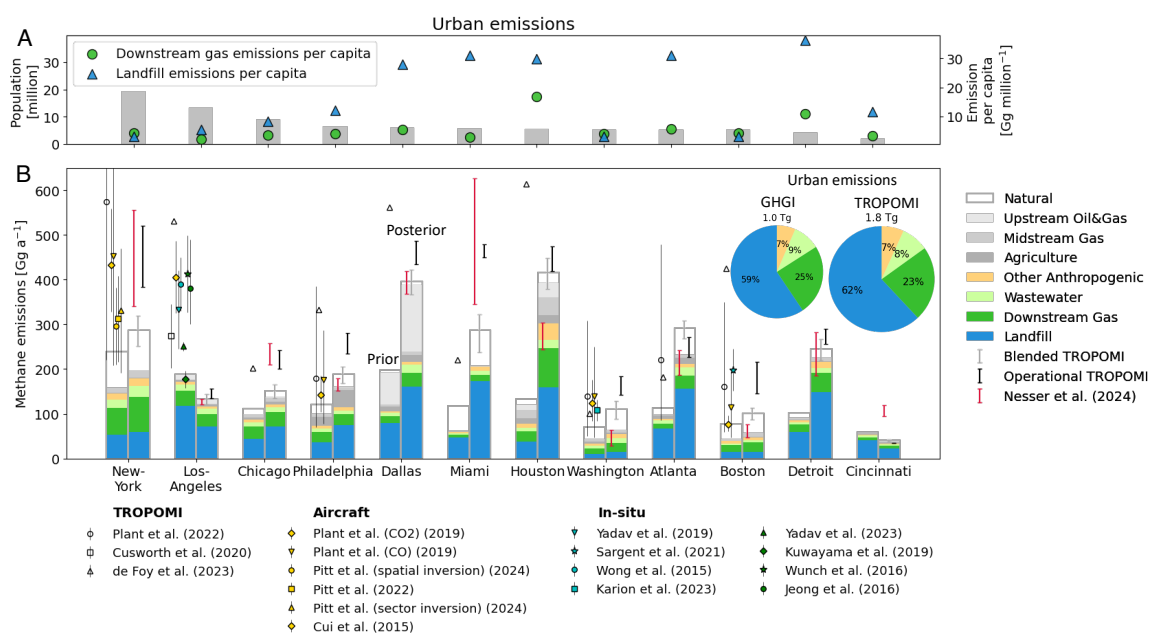

**Fig. 3.** Annual methane emissions in 12 US urban areas ordered by population (largest to smallest). Panel A shows urban population (grey bars) and per capita posterior emissions from our TROPOMI inversion for downstream gas and landfills. Panel B compares our posterior estimates for 2022 subdivided by sectors to the prior estimates from the US EPA GHGI and to previous studies. Vertical bars are error standard deviations of total emissions from the inversion ensemble in individual urban area. Pie charts show total emissions by sector for the 12 urban areas.

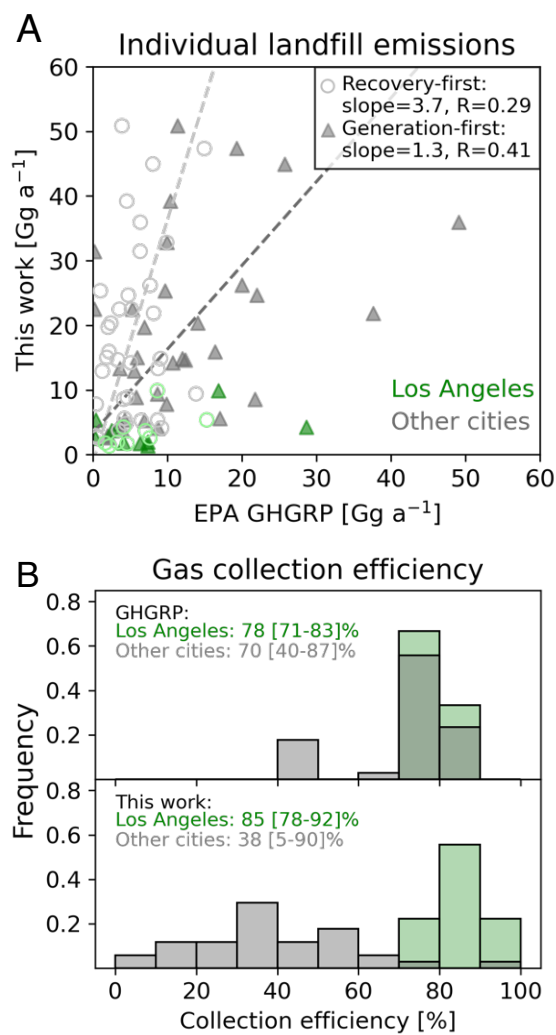

**Fig. 4.** Emissions from 53 urban landfills for which our TROPOMI inversion provides individual information, of which 43 are equipped with gas collection and control systems (GCCS). Panel A compares our posterior estimates for 2022 from the TROPOMI inversion to the GHGRP values using either the recovery-first or the generation-first method. Dashed lines give reduced-major-axis (RMA) regressions excluding Los Angeles, with slopes and correlation coefficients shown inset. Panel B shows the frequency distributions of gas collection efficiency (CE in equation (2)) for 9 landfills in Los Angeles and 44 landfills in other cities, comparing GHGRP estimates to values calculated from our posterior emissions. Inset values give the means and ranges.